\documentclass[amssymb,prl,showpacs,twocolumn,citeautoscript,floatfix,
footinbib,superscriptaddress]{revtex4}
\usepackage{color,graphicx}
\usepackage[latin1]{inputenc}
\usepackage{times}
\usepackage{mathptm}

\begin{document}

\preprint{APS/123-QED}

\title{Thermal conductivity across the metal-insulator transition in single crystalline hyperkagome Na$_{3+x}$Ir$_3$O$_8$}

\author{B. Fauqu\'e}
\affiliation{H. H. Wills Physics Laboratory, University of Bristol, Tyndall Avenue, Bristol, BS8 1TL, United Kingdom}
\affiliation{LPEM (UPMC-CNRS), Ecole Superieure de Physique et de Chimie Industrielles, Rue Vauquelin, 75005 Paris, France}

\author{Xiaofeng Xu}
\affiliation{H. H. Wills Physics Laboratory, University of Bristol, Tyndall Avenue, Bristol, BS8 1TL, United Kingdom}
\affiliation{Department of Physics, Hangzhou Normal University, Hangzhou 310036, China}

\author{A. F. Bangura}
\affiliation{H. H. Wills Physics Laboratory, University of Bristol, Tyndall Avenue, Bristol, BS8 1TL, United Kingdom}
\affiliation{RIKEN (The Institute of Physical and Chemical Research), 2-1 Hirosawa, Wako, Saitama 351-0198, Japan}

\author{E. C. Hunter}
\affiliation{School of Physics, University of Edinburgh, Mayfield Road, Edinburgh EH9 3JZ,UK}

\author{A. Yamamoto}
\affiliation{RIKEN (The Institute of Physical and Chemical Research), 2-1 Hirosawa, Wako, Saitama 351-0198, Japan}

\author{K. Behnia}
\affiliation{LPEM (UPMC-CNRS), Ecole Superieure de Physique et de Chimie Industrielles, Rue Vauquelin, 75005 Paris, France}

\author{A. Carrington}
\affiliation{H. H. Wills Physics Laboratory, University of Bristol, Tyndall Avenue, Bristol, BS8 1TL, United Kingdom}

\author{H. Takagi}
\affiliation{RIKEN (The Institute of Physical and Chemical Research), 2-1 Hirosawa, Wako, Saitama 351-0198, Japan}
\affiliation{Department of Physics, University of Tokyo, Hongo, Tokyo, 113-0033, Japan}

\author{N. E. Hussey}
\affiliation{H. H. Wills Physics Laboratory, University of Bristol, Tyndall Avenue, Bristol, BS8 1TL, United Kingdom}
\affiliation{High Field Magnet Laboratory, Institute for Molecules and Materials, Radboud University Nijmegen, Toernooiveld 7, 6525 ED Nijmegen, The Netherlands}

\author{R. S. Perry}
\affiliation{School of Physics, University of Edinburgh, Mayfield Road, Edinburgh EH9 3JZ,UK}
\affiliation{London Centre for Nanotechnology, 17-19 Gordon Street, University College London, London WC1H 0AH, United Kingdom}

\date{\today}\begin{abstract}
The hyperkagome antiferromagnet Na$_{4}$Ir$_3$O$_8$ represents the first genuine candidate for the realisation of a three-dimensional quantum spin-liquid.
It can also be doped towards a metallic state, thus offering a rare opportunity to explore the nature of the metal-insulator transition in correlated, frustrated magnets. Here we report thermodynamic and transport measurements in both metallic and weakly insulating single crystals down to 150 mK. While in the metallic sample the phonon thermal conductivity ($\kappa^{ph}$) is almost in the boundary scattering regime, in the insulating sample we find a large reduction $\kappa^{ph}$ over a very wide temperature range. This result can be ascribed to the scattering of phonons off nanoscale disorder or off the gapless magnetic excitations that are seen in the low-temperature specific heat. This works highlights the peculiarity of the metal-insulator transition in Na$_{3+x}$Ir$_3$O$_8$ and demonstrates the importance of the coupling between lattice and spin degrees of freedom in the presence of strong spin-orbit coupling.
\end{abstract}

\pacs{Valid PACS appear here}
\maketitle

With their low energy physical response dominated by geometrical frustration, strong electron correlations and fractionalized elementary excitations, quantum spin liquids (QSLs) occupy a special place in the arena of quantum magnetism. While QSLs have received sustained theoretical attention for many years \cite{Anderson1973}, it is only in the last decade that candidate systems have been discovered \cite{Lee2008a, Balents2010}, and only one of these, the hyperkagome Na$_4$Ir$_3$O$_8$, is based on a {\it three-dimensional} (3D) network of corner-sharing triangles. Na$_4$Ir$_3$O$_8$ has a Curie-Weiss temperature $\Theta_{CW}$ = 650K, yet the localized $S = \frac{1}{2}$ moments on the Ir$^{4+}$ ions do not order down to temperatures as low as 2 K \cite{Takagi2007}. The presence of the 5$d^5$ Ir ions ($Z$ = 77) implies that the physics of Na$_4$Ir$_3$O$_8$ may also be heavily influenced by strong spin-orbit coupling, prompting a large number of theoretical works exploring various scenarios to describe its magnetic, Mott insulating ground state \cite{Kim2007, Lawler2008, Lee2008, Balents2008, Balents2008a, Norman2010, Bergholtz2010, Chen2013}.

Specific heat and thermal transport measurements are considered to be insightful, if indirect probes of the ground state of candidate QSLs. Specific heat $C$ reveals the presence of low-lying excitations (including those exclusive to spins), while thermal conductivity $\kappa$ indicates if such excitations propagate long enough to be considered delocalized and reveals the strength of coupling between the spin and the lattice degrees of freedom \cite{Berman}. In some cases, magnetic excitations can contribute significantly to the overall thermal conductivity. For example, in Et$_2$Me$_2$SbPd[dmit$_2$]$_2$ (a 2D  spin-$\frac{1}{2}$ Heisenberg triangular lattice), a residual $T$-linear term was observed on top of the usual $T^3$ phonon contribution \cite{Matsuda10}. Such a residual $T$-linear contribution to $\kappa$ is not expected in an insulator and thus it has been viewed as evidence for mobile, fermionic gapless excitations of {\it magnetic} origin. By contrast, in the 3D classical spin liquid Tb$_2$Ti$_2$O$_7$ and the spinel ZnCr$_2$O$_4$, an extremely low $\kappa$ has been observed \cite{Sun13, Tachabina13}, similar to that found in amorphous solids \cite{Berman}. There, the marked reduction in the {\it phonon} contribution was interpreted as a signature of scattering off strong magnetic fluctuations in the absence of long-range order.

For polycrystalline Na$_4$Ir$_3$O$_8$, a $T^2$ specific heat contribution of magnetic origin was first reported in \cite{Takagi2007}, though more recent milliKelvin experiments on powdered Na$_4$Ir$_3$O$_8$ uncovered an additional $T$-linear term \cite{Gegenwart13}. This latter study also revealed a vanishing $\kappa/T$ as $T \rightarrow$ 0 K. Measurements on polycrystalline or powdered samples however are notoriously susceptible to extrinsic effects, such as impurity contributions in the specific heat or grain boundary scattering of phonons in the low-$T$ thermal conductivity, and it is not clear which of these intriguing observations are intrinsic to Na$_4$Ir$_3$O$_8$.

In order to address this issue, we report here an experimental investigation of the specific heat and thermal conductivity of Na$_{3+x}$Ir$_3$O$_8$ single crystals down to milliKelvin temperatures. The Na deficiency of Na$_4$Ir$_3$O$_8$ is found to lead to a doping-induced insulator-metal transition \cite{Takayama13}, while preserving the chiral hyperkagome lattice of Ir atoms, thereby providing a rare opportunity to explore the Mott insulator-metal transition in a QSL in the presence of a large spin-orbit coupling \cite{Podolsky2009, Podolsky2011}. In insulating Na$_{3+x}$Ir$_3$O$_8$, an anomalously small phonon thermal conductivity is observed  over a very wide temperature range, indicative of a significant decrease in the phonon mean-free-path arising from scattering off uncondensed magnetic excitations or nanoscale disorder. In the metallic sample, we observe that phonons are almost in the boundary scattering regime.

A large number of single crystals were pre-characterized in preparation for these experiments by means of electrical resistivity $\rho(T)$ and d.c. magnetisation measurements performed using a standard 4-probe ac lock-in detection technique and a SQUID magnetometer respectively. Samples displayed either weakly insulating or weakly metallic behavior depending on crystal growth conditions.  The crystal growth and characterization are discussed in the supplementary material, the summary of which is that insulating crystals had $x$ = 0.6 and metallic crystals $x$ = 0.0, similar to that reported recently by Takayama {\it et al.}\cite{Takayama13}.  The thermal conductivity measurements were carried out in two stages on two different crystals located at the extremes of the spectrum in $\rho(T)$ behavior, referred to hereafter by the subscripts  \lq {\it met}' and \lq {\it ins}'. High temperature $\kappa(T)$ measurements (10 K $\leq T \leq$ 300 K) were performed using a heat-pipe technique described in detail elsewhere  \cite{Wakeham2011}, while the low-$T$  measurements were conducted in a dilution refrigerator with a standard one-heater two thermometer set up. The specific heat of the same single crystals from the same batches (hence showing similar resistivity behaviour)  was also measured using a relaxation calorimeter described in Ref.~\cite{Owen2007} and on a Quantum Design PPMS.

\begin{figure}[htbp]
\begin{center}
\includegraphics[angle=0,width=7.5cm]{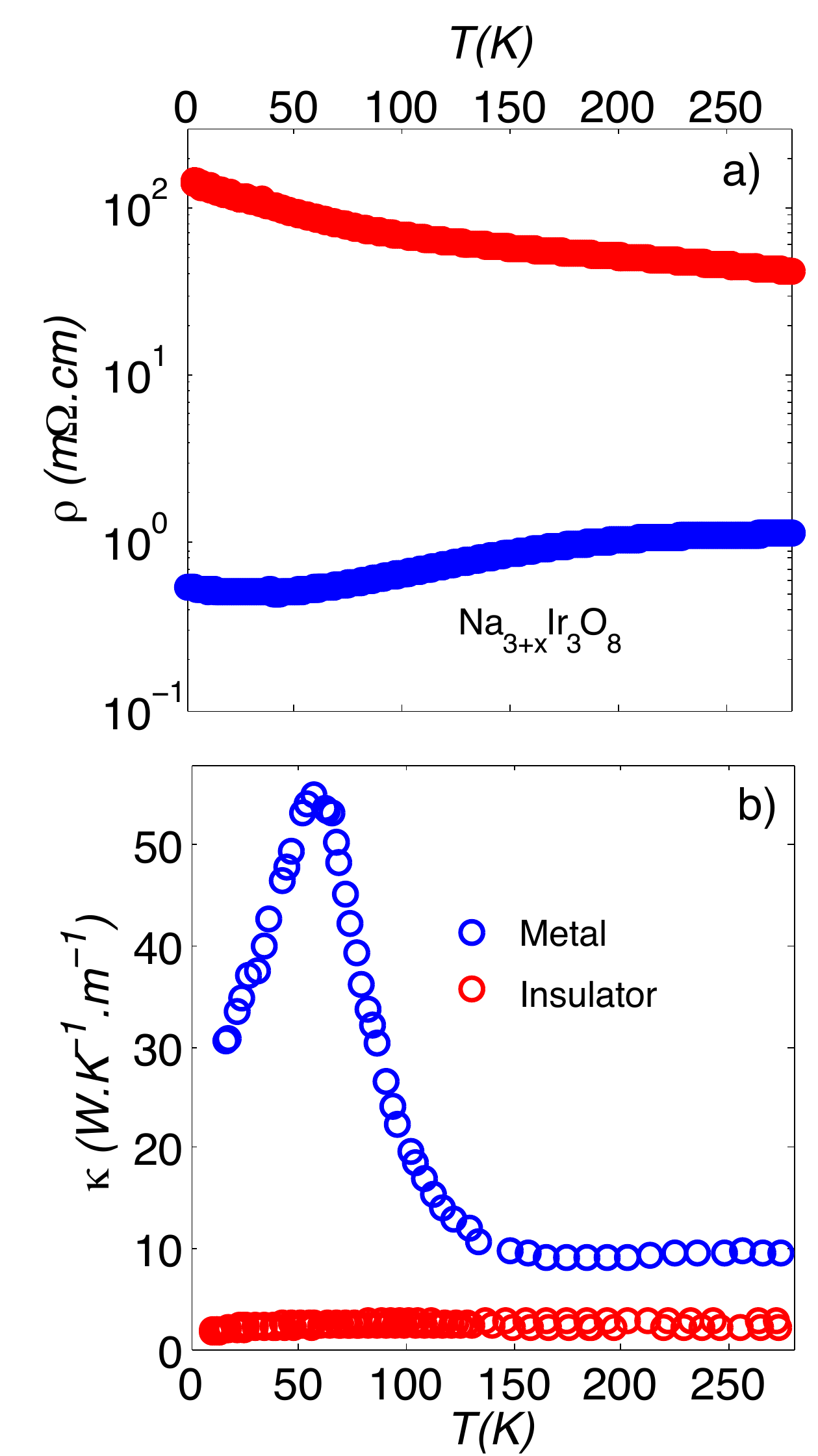}
\caption{{\bf a) Electrical resistivity ($\rho$) of Na$_{3+x}$Ir$_3$O$_8$ single crystals as a function of temperature. b) Thermal conductivity ($\kappa$) of the same samples as a function of temperature. In this and all subsequent figures, the metallike (x=0) and insulating (x=0.6) samples are plotted using blue open and red open circles respectively.}}
\label{Fig1}
\end{center}
\end{figure}

Figure \ref{Fig1} shows the electrical resistivity (panel a) on a semi-log plot) and thermal conductivity (panel b)) as a function of temperature for the insulating sample (red circles) and the metallic sample (blue circles). At room temperature, $\rho_{ins}$ = 40 m$\Omega$cm, rising to 130 m$\Omega$cm at the lowest temperature measured (0.15 K). For the metallic sample, $\rho_{met}$ = 1.1 m$\Omega$cm at room temperature then falls monotonically to $T$ = 25 K, at which point $\rho_{met}(T)$ develops a small upturn. As $T \rightarrow$ 0, $\rho_{met}$ levels off at a value of $\approx$ 600 $\mu\Omega$cm. The high value of $\rho_{met}$ is consistent with the semi-metallic picture for Na$_3$Ir$_3$O$_8$ revealed by recent dc \cite{Takayama13} and optical conductivity measurements \cite{Propper}.

As we can see from Fig.\ref{Fig1}b), both thermal conductivities are temperature independent above $T$ = 150 K with $\kappa_{met} \approx 5\kappa_{ins}$. From the difference in the dc conductivities, it is clear that this is not simply due to the additional electronic term $\kappa^{el}(T)$ in the metallic sample, which, according to the WF ratio ($\kappa^{el}/\sigma T = L_0$ where $L_0 = (\pi^2/3)(k_B/e)^2$), is at least one order of magnitude lower than $\kappa_{met}(T)$. Thus, we can infer that even in the nominally isostructural insulating sample, the phonon thermal conductivity is significantly suppressed over the entire temperature range studied.

Below 100 K, $\kappa_{met}$ develops a strong $T$-dependence, passing through a maximum at $T_{\rm max}$ = 55 K. By contrast, $\kappa_{ins}$ remains essentially constant down to temperatures of order 10 K. The peak observed in $\kappa_{met}$ is the usual behavior expected for the phonon thermal conductivity $\kappa^{ph}$ of a single crystal, reflecting the balance between the decreasing phonon specific heat and the increasing phonon mean-free-path \cite{Berman}. Typically in doped semiconductors, when the carrier density decreases, the phonon mean-free-path increases and likewise, the amplitude of the peak increases \cite{Carruthers57}. In Na$_{3+x}$Ir$_3$O$_8$, we observe the opposite behavior and while insulating Na$_{3.6}$Ir$_3$O$_8$ is expected to be structurally more inhomogeneous (see below), it is clear that we are dealing here with a non-trivial metal-insulator transition.

\begin{figure}[htbp]
\begin{center}
\includegraphics[angle=0,width=7.5cm]{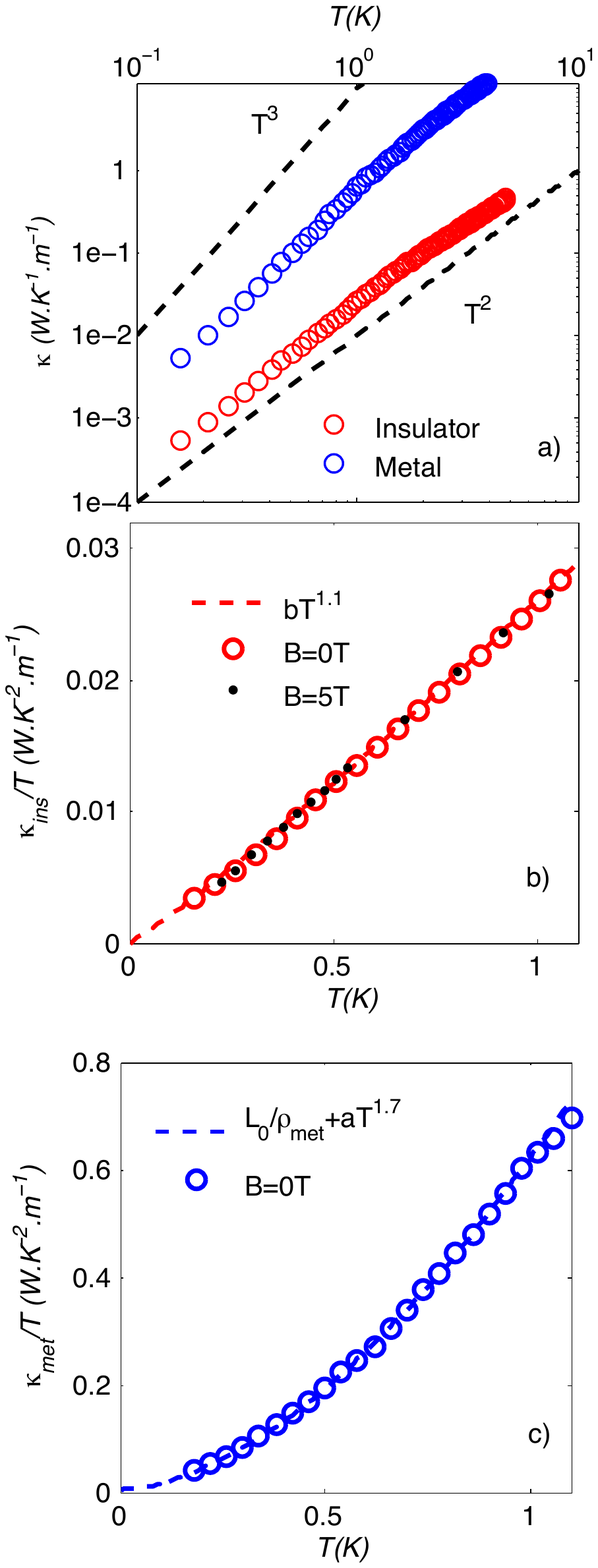}
\caption{{\bf a) Zero-field thermal conductivity of Na$_{3+x}$Ir$_3$O$_8$ between 150 mK and 6 K plotted on a log-log scale b) $\kappa_{ins}/T$ as a function of temperature for the insulating sample (0 Tesla, red open circles; 5 Tesla, black closed circles). The black dashed line is a single-component fit to $\frac{\kappa_{ins}}{T}$ = $b$ $T^\beta$ with $b$ = 0.026 W.K$^{-(2+\beta)}$.m$^{-1}$ and $\beta$ = 1.1. c) $\kappa_{met}/T$ as a function of $T^2$ for the metalike sample (0 Tesla, blue open circles). The black dashed line is a fit to $\frac{\kappa_{met}}{T}$ = $\frac{\kappa^{el}}{T}$+$aT^{\alpha}$ where $\frac{\kappa^{el}}{T}$ = $\frac{L_0}{\rho_{met}}$ = 4.1 $\times 10^{-3}$ W.K$^{-2}$.m$^{-1}$ and $a$=0.61 W.K$^{-(2+\alpha)}$.m$^{-1}$ and $\alpha$ = 1.7.}}
\label{Fig2}
\end{center}
\end{figure}

In order to investigate the origin and evolution of this reduced phonon contribution we extended our measurements to lower temperature.  Figure \ref{Fig2}a) shows a log-log plot of $\kappa_{met}(T)$ and $\kappa_{ins}(T)$ between 150 mK and 6 K. Interestingly, we find not only that $\kappa_{met}$ is significantly enhanced with respect to $\kappa_{ins}$ (the ratio is now $\frac{\kappa_{met}}{\kappa_{ins}} \approx 27$) but also that the temperature dependence of $\kappa_{met}$ and $\kappa_{ins}$ are fundamentally different.The two distinct power-law behaviors are more clearly seen in panels~\ref{Fig2}b) and~\ref{Fig2}c), where we plot respectively $\kappa_{ins,met}/T$ as a function of $T$ together with the best power law fit to the data below 1K:  whereas in the metallic case, $\kappa_{met}$ varies almost as $T^{3}$, $\kappa_{ins} \propto T^{2}$. For both samples, we find that a magnetic field affects neither the magnitude nor the $T$-dependence of the thermal conductivity.

A two component fit was used for the metallic sample: $\frac{\kappa_{met}}{T}$=$\frac{\kappa^{el}}{T}$+$aT^{\alpha}$. The first term corresponds to the electronic part $\kappa^{el}$, the second term corresponds to the phonon contribution. The phonon contribution to $\kappa_{met}$ can be expressed through the kinetic formula: $\kappa^{ph}$ = $\frac{1}{3}C^{ph}\langle v^{ph}\rangle \ell_0$, where C$^{ph}$ = $\beta^{ph} T^{3}$ is the phonon specific heat at low temperature, $\langle v^{ph}\rangle$ is the average phonon velocity and $\ell_0$ is their mean free-path. From the specific heat data discussed below, we find $\beta^{ph}$ = 0.9 mJ.K$^{-4}$.mol$^{-1}$ and $\langle v^{ph}\rangle$ $\approx$ 2800 m.s$^{-1}$, estimated from the Debye temperature ($\Theta_D \approx$  319K) following \cite{Zhao11}. In the ballistic regime, $\ell_0 \approx 2 w/\pi$, where $w$ is the average width of the rectangular-shaped crystal (= 0.027 cm in this case) \cite{Hussey00}. The phonon contribution is then expected to be $\kappa^{ph}$ = $a^{ph}T^3$ where $a$ = 1.4 W.K$^{-4}$.m$^{-1}$. Assuming the WF law is valid for the electronic contribution ($\frac{\kappa^{el}}{T}$ = $\frac{L_0}{\rho_{met}}$=4.1 $\times 10^{-3}$ W.K$^{-2}$.m$^{-1}$), the best fit we find is $a$=0.61 W.K$^{-(1+\alpha)}$.m$^{-1}$ and $\alpha_{met}$=1.7 slightly smaller than the factor two expected in the ballistic regime. The small departure from $T^3$ can either occur if the phonons are not wholly in the boundary scattering regime or when the specular reflections are large such as in a dielectric system like Al$_2$O$_3$ \cite{Pohl82}. In our case, since  $a \leq a^{ph}$, the deviation from $T^3$ is presumably due to the first reason.

\begin{figure}[htbp]
\begin{center}
\includegraphics[angle=0,width=8.5cm]{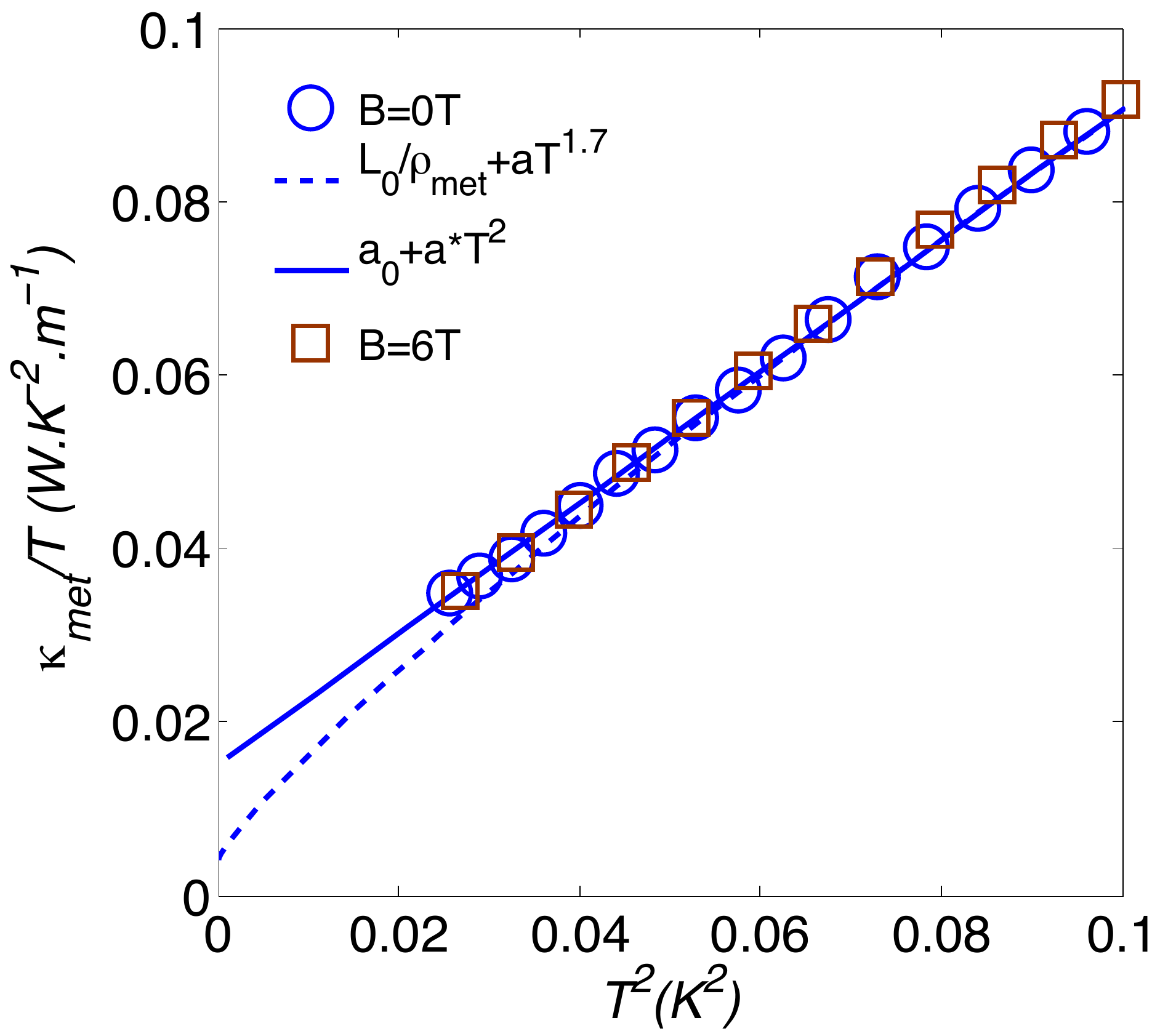}
\caption{{\bf $\kappa_{met}/T$ as a function of $T^2$ for the metallic sample (0 Tesla, blue open circles; 6 Tesla, brown open squares). The blue dash line is a fit to $\frac{\kappa_{met}}{T}$ = $ \frac{L_0}{\rho_{met}}$ +$aT^{\alpha}$ where $ \frac{L_0}{\rho_{met}}$=4.1$ \times 10^{-3}$ W.K$^{-2}$.m$^{-1}$ and $a$ = 0.61 W.K$^{-(2+\alpha)}$.m$^{-1}$ and $\alpha$=1.7. The blue line is a fit to $\frac{\kappa_{met}}{T}$ = $a_0$ + $a^*T^2$ where $a_0 = 1.4 \times 10^{-2}$ W.K$^{-2}$.m$^{-1}$ and $a^*$ = 0.76 W.K$^{-4}$.m$^{-1}$.}}
\label{Fig4}
\end{center}
\end{figure}

As discussed in the context of high-$T_c$ cuprates \cite{Li08,Li09}, the deviation from a purely $T^3$ behavior can significantly affect the estimation of the residual fermionic contribution. In Fig.\ref{Fig4}, we report $\frac{\kappa_{met}}{T}$ as a function of $T^2$ from 300 mK down to 150 mK. The blue line is the fit previously discussed:  $\frac{\kappa_{met}}{T}$=$\frac{L_0}{\rho_{met}}$+$aT^{\alpha}$. The blue dashed line is the textbook fit: $\frac{\kappa_{met}}{T}$= $a_0$ + $a^*$T$^2$ where $a_0$ = 1.4 $\times$ 10$^{-2}$ W.K$^{-2}$.m$^{-1}$ and $a^*$ = 0.76 W.K$^{-4}$.m$^{-1}$. As we can see, the two fits are identical down to 200 mK but the residual intercepts are significantly different. While for the first fit the WF law is verified, for the second fit the residual linear term is three times higher than expected. In the absence of measurements below 150mK, either scenario cannot be ruled out. Yet, such analysis has the merit to quantify the maximum deviation of the WF law. We will come back to this point later.

For the insulating sample, $\kappa(T)$ varies approximately quadratically with temperature from 1 K down to the lowest temperature investigated. The simplest single-component fit of $\kappa_{ins}$ is $\frac{\kappa_{ins}}{T}$ = $b T^\beta$ with $b = 0.026$ W.K$^{-(2+\alpha_{ins})}$.m$^{-1}$ and $\beta = 1.1$. This kind of exponent is reminiscent of that seen in the thermal conductivity of glassy materials where it is attributed to resonant phonon scattering off two-level systems associated with tunneling between closely-spaced atomic configurations \cite{Zeller71}. The absolute value of $\kappa_{ins}$ at $T$ = 1 K is also comparable with that found in amorphous solids such as vitreous silica \cite{Zeller71}.  For such systems, the $T^2$ behavior is associated with a small phonon mean-free-path that varies as  $\ell_0 \approx 1/T$ \cite{Anderson72}. Following this line of thought, we estimate that the magnitude of the phonon mean-free-path in the insulating sample is $\ell_0 \approx 5 \mu m$ at $T$ = 1 K.  This combination of three key observations: the absence of a phonon peak at high temperature, the absence of a $T^3$ term in the low-$T$ thermal conductivity and the low value of  $\kappa_{ins}$ all consistently point to an anomalous reduction of the phonon mean-free-path over the entire temperature range studied. A recent X-ray study of Na$_3$Ir$_3$O$_8$ single crystals \cite{Takayama13} suggested the presence of Na$_4$Ir$_3$O$_8$ inclusions or nano-domains in samples that displayed insulating behavior. Such nanoscale disorder could, in principle, given rise to the amorphous-like thermal transport that we observe in our insulating single crystal. We note, however, that the thermodynamic (magnetic susceptibility and specific heat) data are in good agreement with that observed in polycrystalline Na$_4$Ir$_3$O$_8$ \cite{Takagi2007} (single crystal susceptibility data are shown in the Supplementary Material), and quite distinct from what is found in the metallic sample, suggesting that the insulating behavior of our crystals is intrinsic and is not due simply to poor sample quality.

\begin{figure}[htbp]
\begin{center}
\includegraphics[angle=0,width=6.9cm]{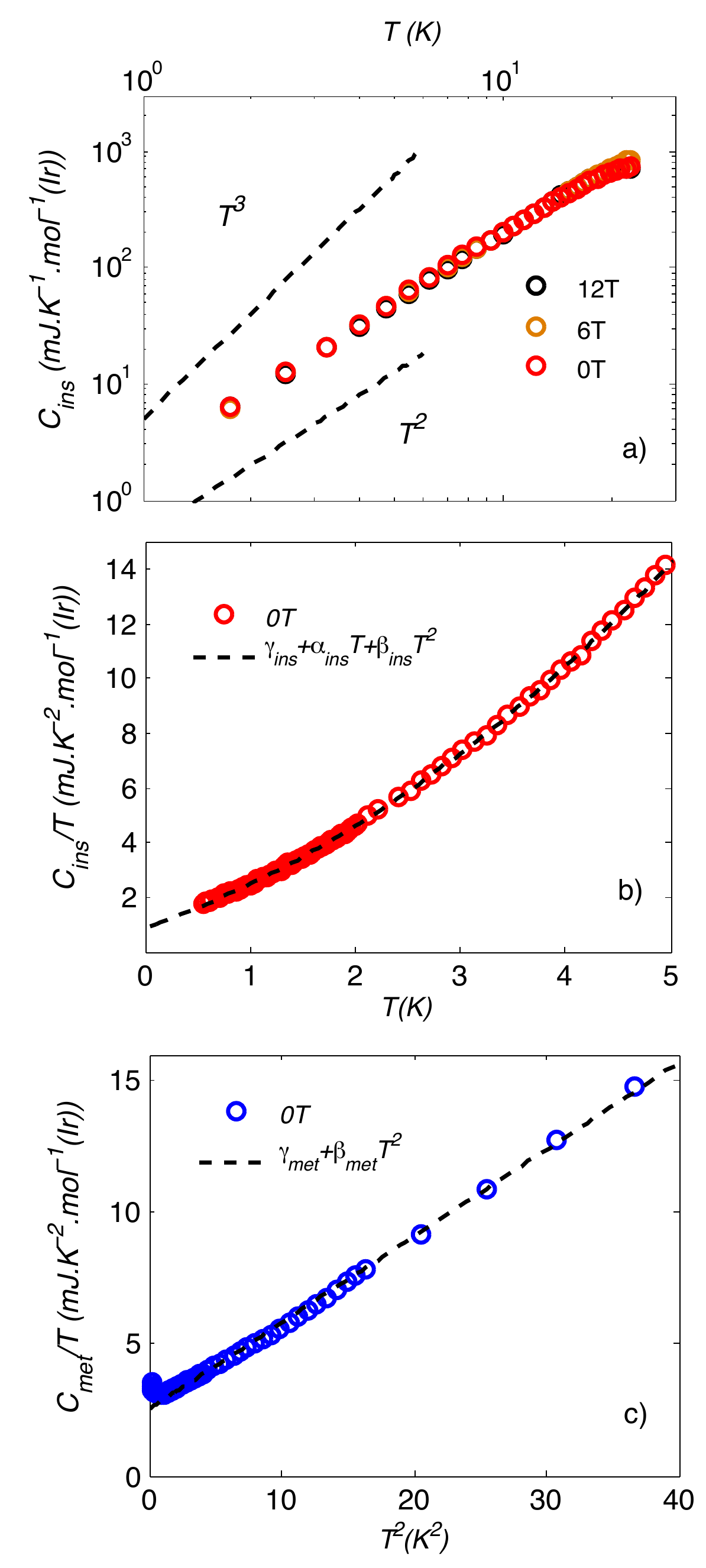}
\caption{{\bf  a) Temperature dependence of the specific heat $C_{ins}(T)$ of insulating Na$_{3+x}$Ir$_3$O$_8$ in magnetic fields of 0 (red), 6 (orange) and 12 Tesla (black open circles) plotted on a log-log scale. b) Low-$T$ magnetic specific heat  of the insulating sample ($C_{ins}$) divided by $T$ as function of $T$. The black line is a fit of  $C_{ins}(T)$ = $\gamma_{ins}T$+ $\alpha_{ins}T^2$ + $\beta_{ins}T^3$ from 0.5 K to 5 K. The values of $\gamma_{ins}$, $\alpha_{ins}$ and $\beta_{ins}$ are given in the text. c) $C_{met}/T$ as a function of $T^{2}$.  The black dotted line is a fit to $C_{met}(T)$ = $\gamma_{met}T$ + $\alpha_{met}T^3$ where $\gamma_{met} $ and $\alpha_{met}$ are given in the text. }}
\label{Fig3}
\end{center}
\end{figure}

These findings are highly reminiscent of the phonon-glass-like behavior recently reported in Tb$_2$Ti$_2$O$_7$ \cite{Sun13,Tachabina13}. There, it was argued that the small energy  separation between the crystal field ground state and the first excited state is the essential ingredient that leads to a strong reduction of $\ell_0$. In particular, the complicated field dependence of the specific heat and thermal conductivity observed in Tb$_2$Ti$_2$O$_7$ has been discussed in the context of spin-splitting of the crystal field levels \cite{Sun13}. In Figure \ref{Fig3}a), we plot the temperature dependence of the specific heat of our insulating sample  for three magnetic field strengths, $B$ = 0, 6 and 12 Tesla. In marked contrast to what is seen in Tb$_2$Ti$_2$O$_{7}$, a magnetic field appears to have no effect on the absolute value of either the specific heat (in good agreement with an earlier powder measurement \cite{Takagi2007}) or the thermal conductivity.

In the light of this finding, a different origin must be invoked to explain the poor thermal conductivity of our weakly insulating Na$_{3+x}$Ir$_3$O$_8$ crystal. A comparison of the specific heat  of the metallic and insulating samples provides a possible clue. As reported in Fig.\ref{Fig3}b) and c), $C_{met}$ and $C_{ins}$ display markedly different behavior at low temperatures. Below 5K, in good agreement with \cite{Takayama13}, $C_{met}(T)$ is well captured by a two component fit $\gamma_{met}T$ + $\beta_{met}T^3$ where $\gamma_{met}$ = 2.4 mJ.K$^{-2}$.mol$^{-1}$(Ir) and $\beta_{met}$ = 0.32 mJ.K$^{-4}$.mol$^{-1}$(Ir). We note that below 1 K a small upturn is observed, presumably due to disorder \cite{Thomas81}. For the insulating sample, as highlighted in Fig.\ref{Fig3}a), the full specific heat is dominated by a $T^2$-term above 1K  with an extra contribution below 1K. Therefore in order to fit across the whole temperature region a third component has to be added to fit $C_{ins}$ = $\gamma_{ins}T$ + $\alpha_{ins}T^2$ + $\beta_{ins}T^{3}$ where $\gamma_{ins}$ = 1.2 mJ.K$^{-2}$.mol$^{-1}$(Ir), $\alpha_{ins}$= 1.0 mJ.K$^{-3}$.mol$^{-1}$(Ir) and $\beta_{ins}$ = 0.31 mJ.K$^{-3}$.mol$^{-1}$(Ir). Note that the $T^3$ terms are almost identical, suggesting that there is no additional $T^3$ term in the insulating sample coming from the presence of two-level-systems \cite{DeYoreo86}. Intriguingly, the  $\gamma$ term does not change significantly across the metal-insulator transition ($\gamma_{met} \approx 2\gamma_{ins} $). It is therefore tempting  to ascribe the $T^2$ term  in the specific heat to the same excitations that are responsible for the strong reduction of the phonon mean-free-path in insulating Na$_{3.6}$Ir$_3$O$_8$. Indeed, this enigmatic $T^2$ term is observed in other kagome lattices \cite{Harry2013,Nakatsuji2005}. It is also noted that the sample with the highest magnetic specific heat  (above 1 K)  has the smallest thermal conductivity. This unexpected finding reveals the complex interplay between the magnetic and phononic degrees of freedom in a system with a large spin-orbit coupling.

One outstanding question of this work is the contribution of the magnetic excitations to the thermal conductivity. Recent investigations on powdered Na$_4$Ir$_3$O$_8$ find an upper limit to the residual linear contribution of $\kappa_{ins}(0)/T$ = 6.3 $\times 10^{-2}$ mWK$^{-2}$m$^{-1}$ \cite{Gegenwart13}. Outside of the ballistic regime, any extrapolation of the phonon contribution, and by inference, of the magnetic contribution, cannot be done reliably.  Nevertheless, it is clear that the magnetic contribution to $\kappa_{ins}$ is extremely small, at least three orders of magnitude smaller than that found in Et$_2$Me$_2$SbPd[dmit$_2]$$_2$ \cite{Matsuda10}. \color{black} In the metallic sample, the ballistic regime is almost attained. A finite linear term can be resolved: it is at maximum, three times larger than what is predicted by the WF law. This deviation could have various origins. It could be associated with the proximity to the metal-insulator transition, as seen for example in underdoped high-$T_c$ cuprates \cite{Proust05} where the deviation has been attributed to intrinsic disorder associated with carrier doping. It could also be associated with the compensated nature of the electronic structure of Na$_3$Ir$_3$O$_8$ \cite{Takayama13}. Here a third term has to be added to $\kappa$ in addition to the phonic and electronic contributions. This third term, the bipolar thermodiffusion, is the result of the heat transport of electron-hole pairs \cite{Gallo62}. In semi-metals such as bismuth, this term can be as high as one quarter of the electronic contribution at high temperature \cite{Gallo63}. This contribution generally collapses at low temperature where the WF law is obeyed \cite{Chau74,Pratt78}. However, the striking similarity of the $\gamma$ terms in both the metallic and insulating samples may indicate another origin for this deviation. While in the insulating sample, the residual specific heat can be attributed to gapless magnetic excitations \cite{Gegenwart13}, an assignment of $\gamma_{met}$ purely to the conduction electrons \cite{Takayama13} is not as clear-cut. Indeed, close to the metal-insulator transition, the distinction between each degree of freedom becomes much less straightforward. Thus, it is feasible that an additional residual term in $\kappa_{met}$ (i.e. above the WF law estimate) could be due to a residue of fermionic magnetic excitations. Roughly speaking, if we assume that $\gamma_{met}$ is due entirely to magnetic excitations (and assuming that the sound velocity of these excitations v$_{mag}\approx $ 10$^5$m.s$^{-1}$), then their maximum mean free path for the magnetic excitation is estimated to be of the order of 10 nm (i.e. $\sim$ 10 unit cells), higher than in the insulating sample, but still two orders of magnitude lower than what has been reported for Et$_2$Me$_2$SbPd[dmit$_2]$$_2$. It would appear then that while the magnetic specific heat is large, the magnetic excitations in Na$_{3+x}$Ir$_3$O$_8$ are either localized (on the insulating side) or have a very short mean free-path (on the metallic side), resulting in a poor thermal conductivity.
\color{black}


In conclusion, by comparing the behavior of two single crystalline samples located on either side of the (albeit weak) metal-insulator boundary, we have uncovered signatures of a significant damping of the phonon mean-free-path in the frustrated hyperkagome antiferromagnet Na$_{3+x}$Ir$_3$O$_8$. This damping may arise from strong coupling of the lattice degrees of freedom with unconventional low-lying spin excitations, possibly mediated through the enhanced spin-orbit coupling of the Ir$^{4+}$ ions. This work has also highlighted the very distinct behavior seen in the specific heat and thermal conductivity of different candidate quantum spin liquids, presenting a significant challenge to those seeking a unified theoretical description. Looking forward, it would be interesting to explore at lower temperature the possible deviation of the WF law across the metal-insulator transition and to investigate whether such strong spin-orbit coupling also influences the low-$T$ thermodynamic and transport properties of other insulating iridates, such as the $j_{\rm eff}$ = $\frac{1}{2}$ Mott insulator Na$_2$IrO$_3$.

\end{document}